# A Game Theoretical Approach for QoS Provisioning in Heterogeneous Networks

A.S.M. Zadid Shifat [1], Mostafa Zaman Chowdhury [1], and Yeong Min Jang [2]

[1] Department of Electrical and Electronic Engineering, Khulna University of Engineering & Technology (KUET), Khulna, Bangladesh (e-mail: zadid_shifat@outlook.com, mzceee@yahoo.com )

[2] Department of Electronics Engineering, Kookmin University, Seoul, Korea (e-mail: yjang@kookmin.ac.kr)

**ABSTRACT**
With the proliferation of mobile phone users, interference management is a big concern in this neoteric years. To cope with this problem along with ensuring better Quality of Service (QoS), femtocell plays an imperious preamble in heterogeneous networks (HetNets) for some of its noteworthy characteristics. In this paper, we propose a game theoretic algorithm along with dynamic channel allocation and hybrid access mechanism with self-organizing power control scheme. With a view to resolving prioritized access issue, the concept of primary and secondary users is applied. Existence of pure strategy Nash equilibrium (NE) has been investigated and come to a perfection that our proposed scheme can be adopted both increasing capacity and increasing revenue of operators considering optimal price for consumers.

## I. INTRODUCTION

With a view to flourishing the coverage and enhancing the capacity as well as data rates in cellular wireless networks, one of the efficient techniques is to alleviate the cell size and transmission distances. Therefore, the concepts of deploying femtocells as well as optical femtocells has recently fascinate growing interests for industrial as well as residential applications. Nevertheless, femtocells and macrocells sharing the same spectrum outcomes in severe interference problems that degrades each other's performance [1]. Cross-polarization allocation concept is already inaugurated in line-of-sight (LOS) systems as a means of sequestrating the interfering signal from the desired signal [2].

In case of femtocell deployments, the prime option is the set of users, those are allowed to access femtocell which are generally known as *subscribers*. Closed access circumscribes the set to specifically registered users, while any mobile *subscribers* are allowed to use any femtocell in open access modes. The preferable criteria strongly depends on the distance between the macro-cellular base station (MBS) and femtocell access points (FAP). Scientists have investigated that residential users prefer closed access whereas the industrial users bring up open access modes [3]. Specific user equipment (UE) those possess proper authorization, i.e., *subscribers*, are allowed to access the corresponding FAP in order to improve their own system throughput, coverage etc. in the closed access mode while the *nonsubscribers* are not allowed to access the closed accessed FAP. Shared access policy has been proposed with a view to alleviating this conflict and solving the issue [4]. As these schemes do not show satisfactory performance in case of Heterogeneous Networks (HetNets), scientists came to light with the applications of game theory and distributed utility functions which are designed in [5] to achieve satisfactory signal to interference plus noise ratio (SINR) for the femtocells. Self-organized power control scheme are proposed in [6] thus *nonsubscribers* can dig up the signals from MBS even if they are located close to FAP.

A cost effective frequency planning for capacity enhancement is proposed in [7] and call admission based adaptive band allocation scheme is proposed in [8] to deal with the interference mitigation strategy and assurance of Quality of Service (QoS). On the basis of co-operation and coalition between UEs, some researchers have proposed co-operative power game as a consolidated solution for the access control problem [9]. Apart from these, game-theoretical model is fabricated in [10] to analyze interaction between the customers and operators. It has been illustrated in [11] that, interference is alleviated and throughput can be improved in open access mode for the entire HetNets under feasible numbers of FAPs and UEs. Moreover, many researchers considered the economic prospects along with the QoS provisioning and suggested a gaming model for different types of access strategy in [12]. In order to reduce co-channel interference, hybrid access mode is applicable as it allows limited number of nonsubscribers to connect to FAPs and the operators may choose open access for increasing their revenues [13]. Definitely, choosing different access policy and different steps for reducing interference as well as increasing capacity are the limitations in present wireless networks.

In our works, we propose dynamic channel allocation strategy based on game theory along with the concept of primary- secondary users. A hybrid access policy is to be adopted along with self-organizing power control scheme. For our proposed scheme there is no need of using different access policy for different purposes. By adopting the proposed scheme, system throughput, capacity as well as revenue of operators can be increased considering optimal price for consumers. From the numerous simulation results, we justified our statements.

The rest of this paper is organized as follows. Proposed model along the mathematical analysis are discussed in Section II. Section III illustrates the performance evaluations along with discussions. Finally, conclusions are drawn in section IV.

## II. PROPOSED MODEL

### A. Dymanic Channel Allocation Scheme

A more effective and economical scheme is proposed here based on dynamic channel allocation scheme. At first, different FAPs are installed and categorized for voice calls and data users which are allocated and embellished by the MBS. FAPs always accumulate feedback information from their environment. Utilizing these information, FAPs allocate obligate channels to the users and change their strategies dynamically to provide better QoS. In this proposed scheme, FAPs are categorized into two prospects such as for voice communication and for data communication. FAPs always count the number of users under itself and if a FAP of a particular category finds that it has less number of users than its capacity, the unused number of channels are then automatically allocated to the other category. It implores for channels to other when it comes into view that it needs more channel than it has already occupied. Supposing that, the asked one have unused channels then they provide channels to the indigent ones. In this prospects, game theoretic algorithm choses the right strategy of a particular user at a particular situation.

### B. Femtocell Access Scheme

In this section, a game theoretical framework based algorithm is established for the HetNets. In this proposed algorithm, each player has the capability of selecting its own strategy. Either the player will target femtocell or choose macrocell as its access network. Let's consider a HetNet scenario containing both macrocell and femtocell in together. Let $P$ and $P^*$, respectively, be defined as a set of User Equipements (UE) whose received signal powers (RSP) from FAP are larger and smaller than that from MBS. We suggest that all *subscribers* are considered in $P$ set in order to access their subscribing femtocell. The *nonsubscribers* can be classified as two types according to whether they are in $P$ or $P^*$ sets. So, the *nonsubscribers* which are in $P^*$ set will have the tendency to select macrocell as their target network and are considered independent as they have no effect on gaming algorithm. On the contrary, in order to explore maximum utilities, *nonsubscribers* which are in $P$ set will have the choice to connect either to MBS or FAP.

The set of *nonsubscribers* $\in P$ are persuaded as the players in the game. It is assumed that *subscribers* can only connect to the FAP. Although *subscribers* are not players in this game, they can affect the game providing sufficient interference. Apart from this, players those connect to MBS and FAP are defined as the macro players and femto players, respectively. Based on dynamic channel allocation scheme, both *nonsubscribers* $\in P^*$ and macro players are considered as Macro User Equipments (MUE) while the *subscribers* and femto players are denoted as Femto User Equipments (FUE).

### C. Analytical Model

Considering the HetNet, a cell selection game is defined as a triplet [13] as follows:

$$\langle I, (s_i)_{i\in 1}, (u_i)_{i\in 1} \rangle \quad (1)$$

where, $I = \{1, 2, 3 ... N\}$ is the set of nonsubscribers $\in P$ i.e., finite set of players. $(s_i)_i \in I$ illustrates the set of pure strategies, $s_i$ and $s_{-i}$ are the non-empty set of actions and opposite reactions, respectively, for player $i$. Therefore, the utility function for each player $i$ can be stated as follows [13]:

$$u_i = (s_i, s_{-i}) = \begin{cases} u_0(0, \|s_{-i}\|_0), & s_i = 0 \\ u_1(1, \|s_{-i}\|_0), & s_i = 1 \end{cases} \quad (2)$$

A round robin scheduling has taken into account for downlink scenario and the time dimension is divided into equal length slots ($T$) for the MUEs. The scheduling in macrocell is done with $X$ nonsubscribers $\in P^*$ and $m$ players connecting to FAPs with $Z-m$ players connecting to MBS which indicates there are total $X+Z-m$ MUEs connecting to MBS where, $Z$ denotes total available players. For each UE, received signal reference powers ($S$) from serving cell and neighboring cell are estimated by this UE and will be reported back to its serving cell [13]. Let $S_{j,a}^{MUE}$ be the received power of 'j' th FUE from BS $a$ ($a = 0$ for MBS; while $a = 1$ represents FAP), which can be written as:

$$S_{j,a}^{MUE} = P_{MAX,a} \cdot L_{j,a}^{MUE} \quad (3)$$

The maximum transmission power of BS '$a$' and large-scale channel gain between BS '$a$' and '$j$'th MUE is represented by $P_{MAX,a}$ and $L_{j,a}^{MUE}$, respectively. In order to reduce coomunication overhead between MBS and FAP during the time slot $T$, allowable power of FAP has two power levels which is as follows:

$$p_j = \begin{cases} P_{MAX,1}, & S_{j,0}^{MUE}[dBm] \geq S_{j,0}^{MUE}[dBm] + \gamma + \psi \\ 0 & S_{j,0}^{MUE}[dBm] < S_{j,0}^{MUE}[dBm] + \gamma + \psi \end{cases} \quad (4)$$

where, the parameter $\gamma$ and $\psi$ are defined as threshold to determine the co-channel, adjacent channel interference and co-tier, cross tier interference level respectively. Larger the values of $\gamma$ and $\psi$, lesser the values of interference. The cell capacity of FAP (considering m players in action) is assumed as:

$$C_{FAP}(m) = \frac{W}{X+Z-m} \sum_{j=1}^{X+Z-m} \log_2 \left(1 + \frac{p_k \cdot L_{W,1}}{\xi^2 Z + P_{MAX,0} \cdot L_{W,0}}\right) \quad (5)$$

In (5), $W$, $L_{W,a}$, and $\xi^2 Z$, respectively, represent system bandwidth, large-scale channel gain between BS '$a$' and worst player, and additive white Gaussian noise (AWGN) power respectively. When macro players utilize common slots, FAP has no privilege to transmit any power during these slots. If FAP transmits during those time slots, macro players will receive strong interference signal due to the reason that $S_{k,1}^{MUE} + \gamma + \zeta > S_{k,0}^{MUE}$ which is valid for $Z + X - n \geq j \geq X + 1$

Let, $D = \sum_{j=1}^{X} \amalg \left(S_{j,0}^{MUE} + \gamma + \zeta > S_{j,0}^{MUE}\right) \quad (6)$

For this type of equation $\amalg$ (True) = 1 and $\amalg$ (False) = 0 [13]. So (5) can be modified as:

$$C_{FBS}(m) = \frac{(X-D)W}{X+Z-m} \log_2 \left(1 + \frac{P_{MAX} \cdot L_{W,1}}{\xi^2 Z + P_{MAX,0} \cdot L_{W,0}}\right) \quad (7)$$

Players connecting to FAP are designed with the less scheduling priority as *subscribers* as the *subscribers* empower higher priority to access the FAP. Thereafter, the concept of dynamic priority assigning for primary and secondary users arise here. A system admission control parameter $\beta$, is also used in this paper likewise [13]. It illustrates that all *subscribers* can initially be allocated with the ratio $\beta$ of total bandwidth available for the femtocell and all the FUEs will share the remaining ($1-\beta$) of the resource. For $m$ players, *subscribers* can share $\left(\frac{BZ+Q}{Q^2+Z}\right)$ of available resource where $\beta = 1$ indicates closed access modes following $\beta = 0$ indicates that the femtocell is completely open. However, FAPs can still control the total number of its serving UEs via the value of closed rate $\beta$. Each femto player can be allocated with $\left(\frac{1-\beta}{X+Q+m}\right)$ of the total resource. The utility function $u_i(s)$ for each player $i$ can be written as:

$$u_i(s) = \begin{cases} u_0(\|s\|_0) = \frac{W}{X+Z-\|s\|_0} \log_2 \left(1 + \frac{P_{MAX,0} \cdot L_{W,0}}{\zeta^2 Z}\right), & s_i = 0; \\ u_1(\|s\|_0) = \frac{1-\beta}{Q+\|s\|_0} \cdot C_{FAP}(\|s\|_0) - \chi \cdot \phi \cdot \Gamma, & s_i = 1; \end{cases} \quad (8)$$

where, $\phi$ is defined as the price that the consumers to be charged by the operators. $\chi$ is used here as a constant of 2 Mbps and $\Gamma$ as a periodic adjustor. $\chi \Gamma \|s\|_0$ is the revenue that the operators can receive from the consumers. NE is observed for the condition $0 \leq m \leq Z$.

### III. PERFORMANCE EVALUATION

For performance analysis, we assumed the radius of MBS and FAP as 1 km and 15 m, respectively. Room dimensions where FAP is located is assumed as 20 m × 20 m. Numeric value of path loss, shadowing deviation, wall penetration and the other required parameters are summarized in Table 1.

Table 1. System Parameters

| Parameter | Value |
|---|---|
| Max power of MBS | 40 dBm |
| Max power of FAP | 10 dBm |
| System bandwidth | 6 MHz |
| Number of sub-channels | 30 |
| Number of indoor *subscribers* | 6 |
| Number of indoor *nonsubscribers* | 10 |
| Number of outdoor *nonsubscribers* | 12 |
| Channel noise density | -180 dBm/ Hz |
| Noise figure | 10 dB |
| Wall penetration loss | 15 dB |
| Shadowing deviation | 8 dB |
| Free space propagation loss | 10 dB |
| Distance between MBS and FAP | 500 m |
| Interference threshold | 10 dB |

Average capacity of each UE at NE versus the call admission rate parameter $\beta$ for different types of UEs are illustrated in Fig. 1 where "player", "sub", "nonsub", "system" are denoted as each player, *subscriber*, *nonsubscribers* $\in P^*$ and each UE in the HetNet system, respectively. It is observed that, better average system capacity can be attained in open access mode. Furthermore, at $\beta = 0.8$, the average capacity of *subscribers* outcomes a peak value which emphasizes that, there exists a particular value of $\beta$ for which the performance of *subscribers* in our proposed scheme is better to the conventional schemes.

Fig. 2 illustrates average capacity of *subscriber* and system UE at NE versus the distance $d$ between MBS and FAP in open ($\beta = 0$), hybrid (with $\beta = 0.5$), and closed ($\beta = 1$) access modes. It can be seen that our proposed scheme can achieve better balance of performance between *subscribers* and system UEs compared to open, closed, and conventional hybrid access modes. Fig. 3 illustrates average revenue that operator can charge versus the price in different access modes. It is observed that average revenues in open access mode is larger than that in hybrid access mode and the difference is quite staring [13]. The difference of the average revenues between our proposed scheme and conventional hybrid access policy is marginal.

Peak values in Fig. 3 indicates an optimal price that operators can charge from players for the purpose of maximizing their revenue. These results clearly emphasizes to adopt our proposed scheme as an optimum solution between operators and consumers.

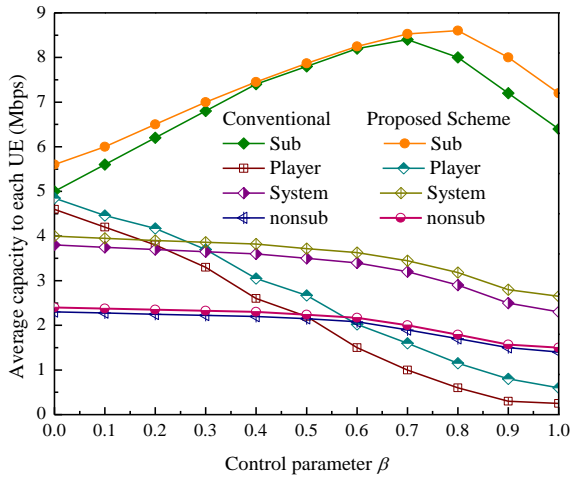

**Fig. 1.** Average capacity of each UE at NE versus $\beta$ for player, *subscriber*, *nonsubscriber* $\in P^*$ and each UE in the system.

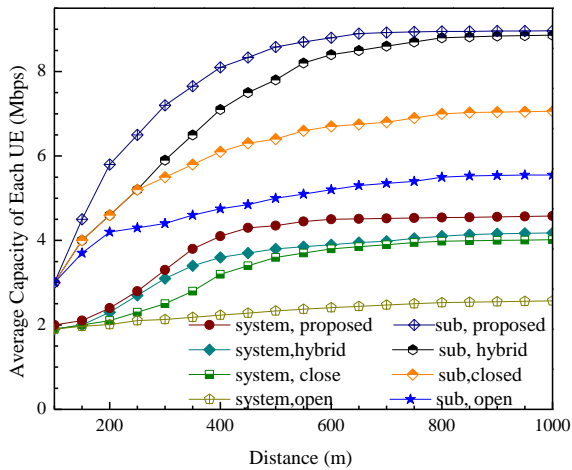

**Fig. 2.** Average capacity of *subscriber* and system UE at NE versus the distance between MBS and FAP.

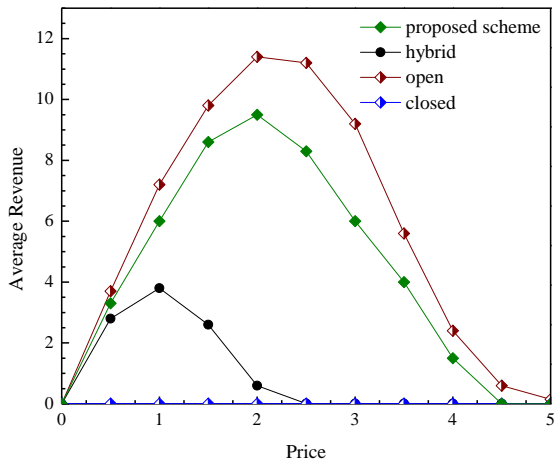

**Fig. 3.** Average revenue (operator) versus the price.

## III. CONCLUSION

In this paper, an efficient game theoretic algorithm is proposed in which dynamic channel allocation has got the superlative priority. From the discussions and graphical illustrations, we have come to a conclusion that, our proposed scheme can be a surveyor to future wireless networks. Although, hybrid access and open access mode can outcome in higher capacity and better revenue for operator, respectively, our proposed scheme can be adopted for increasing revenue with an optimal price for consumer as well as obtaining higher capacity.